\documentclass[preprint,10pt]{JHEP3} 

\JHEPspecialurl{http://jhep.sissa.it/JOURNAL/JHEP3.tar.gz}

\usepackage{epsfig,multicol,amsmath}

\def\beq{\begin{equation}}
\def\eeq{\end{equation}}
\def\bea{\begin{eqnarray}}
\def\eea{\end{eqnarray}}

\def\eq#1{{Eq.~(\ref{#1})}}
\def\fig#1{{Fig.~\ref{#1}}}
\newcommand{\bas}{\bar{\alpha}_S}
\newcommand{\as}{\alpha_S}

\newcommand{\Lb}{\left(}
\newcommand{\Rb}{\right)}

\parskip=\itemsep               

\def\thefootnote{\fnsymbol{footnote}}

\title{\LARGE \bf The  BFKL Pomeron Calculus in zero transverse dimensions:
summation of  Pomeron loops and  generating functional for the
multiparticle production processes }
\author{\large  E. ~Levin\thanks{Email: leving@post.tau.ac.il,
levin@mail.desy.de;} \,\,\,\,and\,\,\,A.~Prygarin \thanks{Email:prygarin@post.tau.ac.il} \\
Department of Particle Physics, School of Physics and Astronomy\\
Raymond and Beverly Sackler Faculty of Exact Sciences\\  Tel Aviv
University, Tel Aviv 69978, Israel}

\abstract{ In this paper we address two problems in the BFKL Pomeron
calculus in zero transverse dimensions: the summation of the Pomeron
loops and the calculation of the processes of  multiparticle
 generation.  We introduce a new generating functional for these processes and obtain the evolution
equation for it. We argue that in the kinematic range given by
 $ 1\,\,\ll\,\,\ln(1/\as^2)\,\,\ll\,\,\,\as\,Y\,\,\ll\,\,\frac{1}{\as}$ ,
  we can reduce the Pomeron calculus to exchange of non-interacting Pomerons with the renormalized
amplitude of their interaction with the target. Therefore, the summation of the Pomeron loops can be
 performed using Mueller, Patel,  Salam and Iancu approximation.}

 \keywords{BFKL Pomeron,  Generating functional,  Mean field approach, Exact solution}

\preprint{  TAUP -2845-07\\
hep-ph/0701178\\
\today}
\begin{document}

\def\thefootnote{\arabic{footnote}}
\section{Introduction}
\label{sec:Int} The problem of  Pomeron interaction in zero
transverse dimensions have been discussed in the framework of the
Reggeon calculus \cite{AMCP} about three decades ago. However,
recently, we have seen a revival of the interest to this problem
(see   \cite{L4,BORY,SHX,KLTM,SHX1,BIT,BMMSX,KKLM} and references
therein). The very reason for this in our opinion is related to a
hope to solve the old problem of finding the high energy asymptotic
behaviour of the scattering amplitude in QCD . We hope for a
solution not in the mean field approximation, where the  solution
has been discussed and well understood both analytically  and
numerically (see  \cite{GLR,MUQI,MV,BK,JIMWLK,LT,IIM,NS}), but in
the approach where the so called Pomeron loops should be taken into
account \cite{MSHW,LELU,IT,KOLU,HIMST}.  The problem of taking into
account the Pomeron loops can be reduced to  the BFKL Pomeron
calculus \cite{BFKL,BART,BRN} or/and to the  solution of statistical
physics problem: Langevin equation and directed percolation
\cite{STPH,EGM,KLP}. The last approach is based on the probabilistic
interpretation of the Pomeron calculus which also has roots in the
past \cite{GRPO,BOR}.

The BFKL Pomeron calculus in zero transverse dimensions being an
oversimplified model has the same description in terms of the
directed percolation as the general approach. Thus  solving this
model we can gain an experience that will be useful for the solution
to a general problem of interaction of the BFKL Pomerons   in QCD.

It is well known that the BFKL calculus in zero transverse
dimensions can be treated as a system that evolves in imaginary time
$i t =  Y$ with the Hamiltonian: \beq \label{I1}
H\,\,=\,\,-\,\Delta\,\phi\,\phi^+\,\,+\,\,\lambda\,\Lb \phi\,\phi^{+
2}\,\,-\,\,\phi^2\,\phi^+\Rb \,\,\,\,\mbox{and   evolution equation
for  wave function}\,\frac{d\,\Psi}{d\,Y}\,\,=\,\,- H\,\Psi \eeq
where  the Pomeron intercept $\Delta\,\,\propto\,\,\as$ and the
triple Pomeron vertex $\lambda \,\,\propto\,\as^2$.

In the next section we will discuss the evolution equation for the
generating functional that describes the system of Pomerons in terms
of probabilities to find `wee' partons (color dipoles \cite{MUCD}).
We  introduce $\Gamma(1 \to 2)\,=\,\Delta$ and $\Gamma(2 \to 1)
\,=\, \Delta\,\gamma$ with   $\gamma$ is the amplitude for low
energy interaction of the colour dipole with the target
(target-Pomeron vertex) and  $\gamma \propto \as^2$. The estimates
for $\Delta$ and $\lambda$ are given in the leading order of
perturbative QCD. We can trust the approach with the Hamiltonian of
\eq{I1} only in the kinematic region of $Y$ given by the following
equation \beq \label{KR}
1\,\,\ll\,\,\ln(1/\as^2)\,\,\ll\,\,\,\as\,Y\,\,\ll\,\,\frac{1}{\as}
\eeq Indeed,  the $n$ Pomeron exchanges give contribution which is
proportional to $\Lb \gamma\,e^{\Delta\,Y} \Rb^2$ and, therefore we
need to sum them in the kinematic region where
$\gamma\,e^{\Delta\,Y} \,\geq\,1$ or $ \Delta\, Y\,\geq\,\ln
(1/\gamma) \approx\,\ln (1/\as^2)$. However, we cannot go to ultra
high energies since we do not know the higher order corrections to
the BFKL kernel and to triple Pomeron vertex. The contribution of
the BFKL Pomeron exchange can be written as $\gamma  e^{(\Delta
\,+\,Const\,\as^2)\,Y}$  and the high order correction term is
essential for $\as\,Y\,>\,\,1/\as$.

In this paper for the kinematic range given by \eq{KR} we obtain two
results. We introduce a new generating functional which allows us to
calculate  processes of multiparticle  generation since it gives us
the probability to have a given number cut and uncut Pomerons (see
\cite{AGK}). We derive the evolution equation for this generating
functional both in the mean field approach and in the approach that
takes into account the Pomeron loops.

The second result is related to the method of summation of the Pomeron loops.
 We claim that in kinematic region of \eq{KR} the Pomeron interaction given by
 Hamiltonian of \eq{I1} can be reduce to the system of free Pomerons with the
 renormalized amplitude of dipole-target interaction at low energies. In other words,
  we can  view the evolution of our system of `wee' partons (
colour dipoles) as a system of not interacting partons only with
emission absorbed in the evolution of the BFKL Pomerons, and all
specific features of this system being determined by the low energy
amplitude of  `wee' parton interaction with the target. Having this
in
 mind we state  that Mueller, Patel, Salam and Iancu \cite{MPSI} approach
gives the solution to the problem.

The paper is organized as follows. In the next section we introduce the mean field approach and discuss
it in the framework of the generating functional. We introduce a new generating functional which gives
 us a possibility to calculate the probability to find a given number of cut and uncut Pomerons. Therefore,
  knowing this generating functional we can calculate the cross section with given multiplicities.
  We derive the evolution equation for this functional. In section three we take into account the Pomeron
  loops and generalize the evolution equation.
In this section we reconsider the problem of summation of all Pomeron loops in the kinematic region of \eq{KR}
and argue that we can reduce this problem to consideration of a system of non-interacting Pomerons with
renormalized vertex of Pomeron-target interaction. Based on this idea we use the Mueller, Patel, Salam and Iancu approach
to calculate
scattering amplitude at high energies  both for elastic and inelastic interactions
with different multiplicities of particles in the final state.

In conclusions we summarize the results and discuss  open problems.

\section{The mean field approximation}
\label{sec:MFA}
\subsection{General approach}

\FIGURE[ht]{
\begin{minipage}{65mm}{
\centerline{\epsfig{file=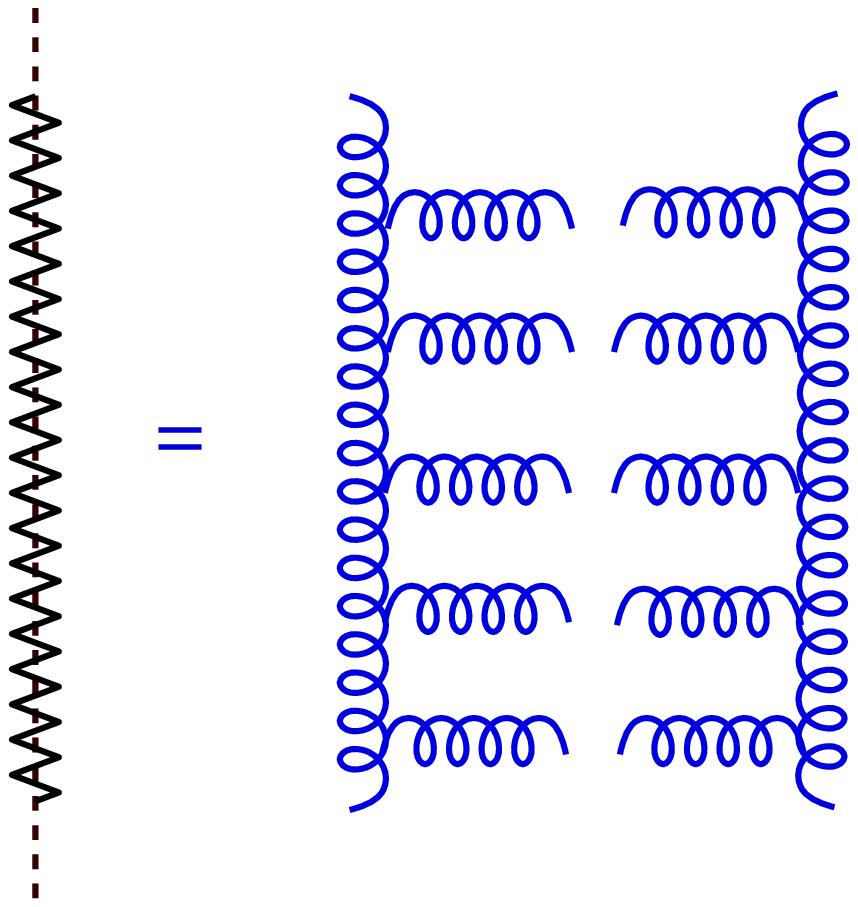,width=60mm}}
}
\end{minipage}
\caption{The diagram for the cut BFKL Pomeron that illustrates \eq{UCP}.  }
\label{cutp}
}

Our approach to multiparticle production is based on the AGK cutting
rules \cite{AGK}. These rules stem from  the unitarity constraint in
$s$-channel, namely, \beq \label{UCP}
2\,Im\,A^{BFKL}(s,b)\,\,=\,\,2\,\,N^{BFKL}(s,b)\,\,=\,\,G^{BFKL}_{in}(s,b)
\eeq where $Im\,A^{BFKL}(s,b) \equiv\,N^{BFKL}(s,b)$ denotes the
imaginary part of the  elastic scattering amplitude for
dipole-dipole interaction at energy $W= \sqrt{s}$ and at the impact
parameter $b$. It is  normalized in the way that  the total cross
section is equal to
 $\sigma_{tot}\,\,=\,\,2\,\int \,d^2b\,\,N^{BFKL}(s,b)$.
 $G^{BFKL}_{in}$ is the contribution of all inelastic  processes for
  the BFKL Pomeron and $\sigma_{in} =\int \,d^2b\,\,G_{in}(s,b)$. Therefore,
   \eq{UCP} gives us the structure of the BFKL Pomeron exchange through the
    inelastic processes and it can be formulated as the statement that the
     exchange of the BFKL Pomeron is related to the processes of multi-gluon
     production in a certain kinematics (see \fig{cutp}). \eq{UCP} is proven in
        \cite{BFKL} in QCD.  Using  it, we can express all processes  of
         multiparticle production  in terms of exchange and interactions  of
 the BFKL Pomerons and/or the cut BFKL Pomerons (see \fig{cutps}).

\FIGURE{
\centerline{\epsfig{file=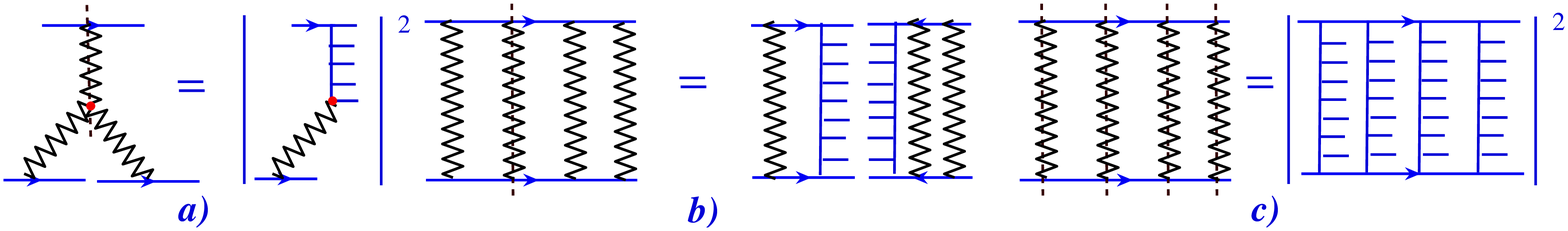,width=160mm,height=30mm}}
\caption{ Several examples of the Pomeron diagrams that contribute
to the multiparticle production:
 the diffraction production of the bunch of particles in the region of rapidity
$\ln(M^2/m^2)\,=\,Y - Y_0$ (\fig{cutps}-a); the process of
multiparticle  production with the average multiplicity due to
exchange of four Pomerons (\fig{cutps}-b); and the process of
multiparticle  production with
 the multiplicity in four times larger than the average multiplicity.}
\label{cutps}
}

The AGK cutting rules establish the relation between different
processes that stems from BFKL Pomeron diagrams. For example,  the
simple triple Pomeron diagrams in \fig{agk3p} leads to three
inelastic processes: the diffractive production of the system with
mass $\ln(M^2/m^2)\,=\,Y - Y_0$ (\fig{agk3p} -A); the multi-gluon
production in the entire kinematic region of rapidity $Y - 0$  with
the same multiplicity of gluons as in one Pomeron ( \fig{agk3p} -B);
and the   multi-gluon production in the region $Y -0$ but with the
same multiplicity of gluons as in one Pomeron only in the rapidity
window $Y - Y_0$ while for the rapidity $Y_0 - 0$ the gluon
multiplicity in two times larger than for one Pomeron( \fig{agk3p}
-C). The AGK cutting rules \cite{AGK} say that the cross sections of
these three processes are related as \beq \label{AGK3P} \sigma_A
\,\,:\,\,\sigma_B
\,\,:\,\,\sigma_C\,\,\,\,\,=\,\,\,\,1\,\,:\,\,-4\,\,:\,\,2 \eeq At
first sight the cross section of the process B is negative but it
should be stressed that one Pomeron also contributes to the same
process and the resulting contribution is always positive.

\FIGURE{ \centerline{\epsfig{file=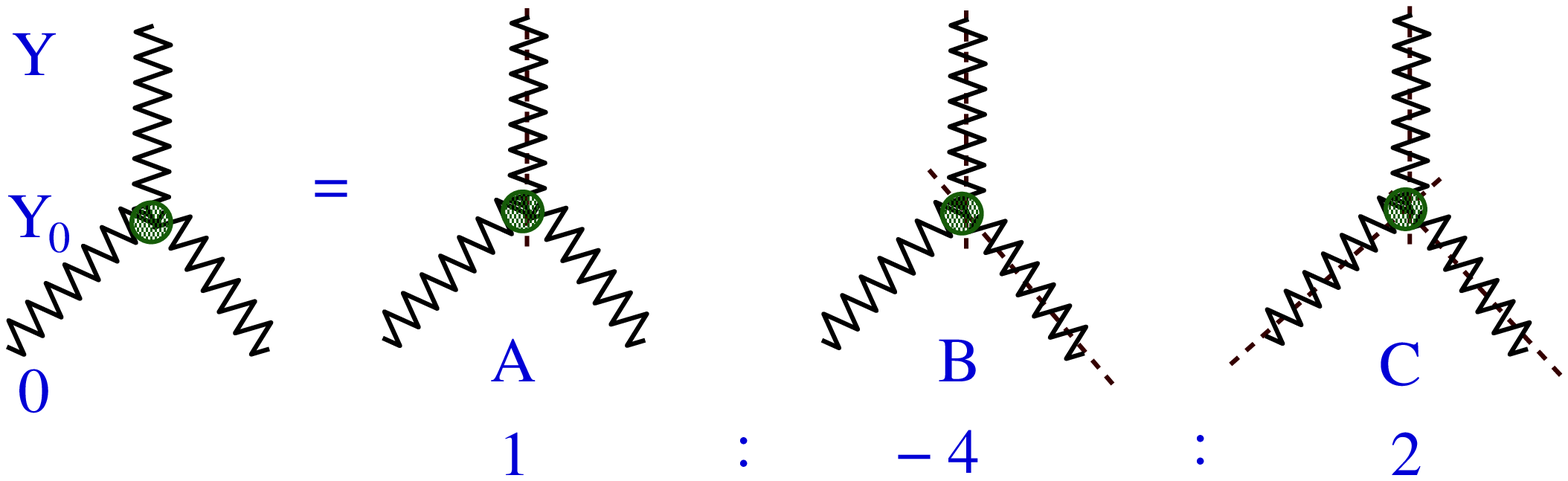,width=140mm}}
\caption{ The AGK cutting rules for the triple Pomeron diagram: the
diffractive production of the system with mass $\ln(M^2/m^2)\,=\,Y -
Y_0$ (\fig{agk3p} -A); the multi-gluon production in the entire
kinematic region of rapidity $Y - 0$  with the same multiplicity of
gluons as in one Pomeron ( \fig{agk3p} -B); and the   multi-gluon
production in the region $Y -0$ but with the same multiplicity of
gluons as in one Pomeron only in the rapidity window $Y - Y_0$ while
for the rapidity $Y_0 - 0$ the gluon  multiplicity in two times
larger than for one Pomeron( \fig{agk3p} -C) } \label{agk3p} }

\newpage

\fig{cutp}  and \fig{agk3p} as well as \eq{AGK3P} allow us to
understand the equation for the single diffractive production in the
mean field approximation that  has been written by Kovchegov and
Levin \cite{KL} and that has the following form (see \fig{kleq} for
graphical representation of this equation) \FIGURE{
\centerline{\epsfig{file=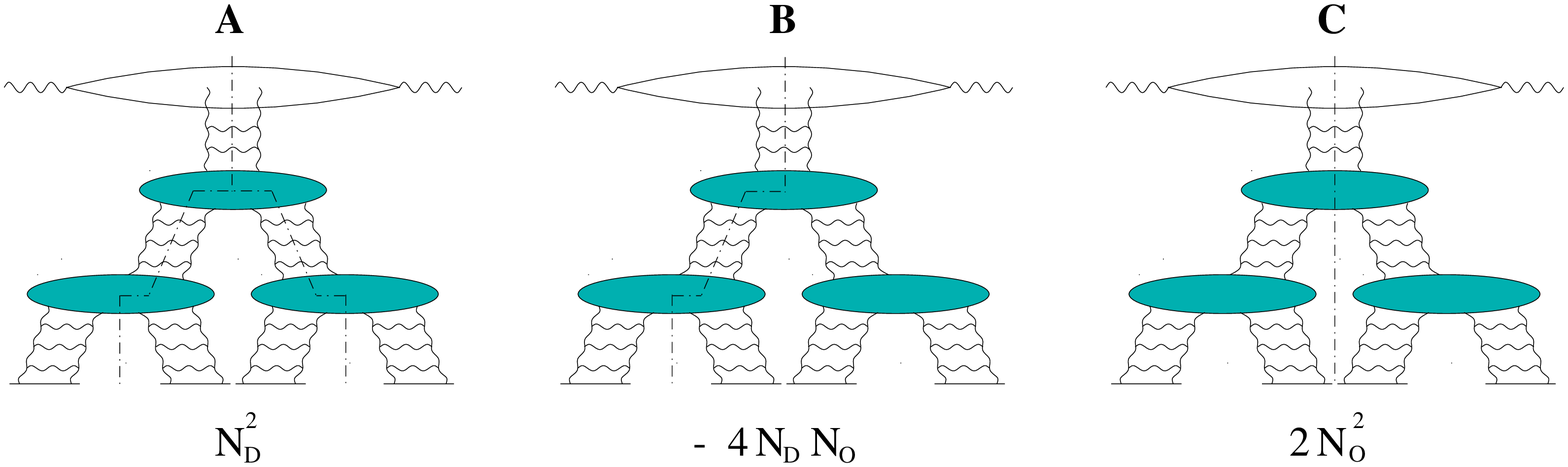,width=140mm}} \caption{ Different
Pomeron cuts  contributing to the cross section of diffractive
dissociation which lead to different terms on the right hand side in
\eq{KLND}.} \label{kleq} }

\beq \label{KLND}
  \dfrac{\partial N^D ({\bf x}_{01},{\bf b}, Y, Y_0)}{\partial Y} =
  \dfrac{ \alpha C_F}{\pi^2} \, \int_\rho d^2 x_2 \,\,
  \dfrac{x^2_{01}}{x^2_{02} x^2_{12}}
\eeq
$$
\left(\, N^D ({\bf x}_{02},{\bf b} + \frac{1}{2} {\bf x}_{12}, Y, Y_0)\,\,+\,\,N^D ({\bf
x}_{12},{\bf b} + \frac{1}{2} {\bf x}_{02}, Y, Y_0)\,\,-\,\,N^D ({\bf x}_{01},{\bf b}, Y,
Y_0) \right.
$$
$$
\left. +\, N^D ({\bf x}_{02},{\bf b} +
  \frac{1}{2} {\bf x}_{12}, Y, Y_0) \, N^D ({\bf x}_{12},
  \frac{1}{2} {\bf x}_{02}, Y, Y_0) \,\,\,\,\,\,\,\,\,\,\,\,\,\, \,\,\,\,\,\,\,\,\,\,\,\,\,\,\,\,\,\,\mbox{ \eq{KLND}(A)}
\right.
$$
$$
\left.- 4 \, N^D ({\bf x}_{02},{\bf b} + \frac{1}{2} {\bf x}_{12}, Y, Y_0)
\, N_0 ({\bf x}_{12},{\bf b} + \frac{1}{2} {\bf x}_{02}, Y) \,\,\,\,\,\,\,\,\,\,\,\,\,\, \,\,\,\,\,\,\,\,\,\,\,\,\,\,
\mbox{\eq{KLND}(B)} \right.$$
$$
\left. + 2 \, N_0
({\bf x}_{02},{\bf b} + \frac{1}{2} {\bf x}_{12}, Y) \, N_0 ({\bf
x}_{12},{\bf b} + \frac{1}{2} {\bf x}_{02}, Y) \right)  \,\,\,\,\,\,\,\,\,\,\,\,\,\,\,\,\,\, \,\,\,\,\,\,\,\,\,\,\,\,\,\,\,\,\,\,
\mbox{\eq{KLND}(C)}
$$

with the initial condition given by \beq \label{INCSD} N^D ({\bf
x}_\perp,{\bf b}, Y=Y_0, Y_0) = N_0^2 ({\bf x}_\perp,{\bf b}, Y_0).
\eeq where $N_0$ is the solution of the Balitsky - Kovchegov
equation \cite{BK} and $N_D(\bf{x},\bf{b};Y;Y_0)$ is the diffraction
dissociation of the colourless dipole with size $x$ at impact
parameter $b$ into the system of gluon with the rapidity gap larger
than $Y_0$ at energy $Y$. At first sight, \eq{KLND}  contradicts the
AGK relations given by \eq{AGK3P} (see \fig{agk3p}), but this
contradiction can be easily resolved if we take into account
coefficient 2 in \eq{UCP} (see   \cite{KL} for more details as well
as for a proof based directly on the dipole approach to high energy
scattering).

Despite a simple structure of \eq{KLND}, which is only a little bit
more complicated than the Balitsky - Kovchegov equation, as far as
we know there exists no analytical solution to this equation and
there is the  only attempt to solve it numerically \cite{NSD}.
However, this equation has a simple solution in the toy model (see
\cite{KL,BGLM,BORY} )   which we are going to discuss.
\subsection{The BFKL Pomeron calculus in zero transverse dimensions: general approach}
The mean field approximation looks extremely simple  in the BFKL
Pomeron calculus in zero transverse dimensions (the toy model
\cite{MUCD,L1,L2}). Indeed, in  the toy model, in which there is no
dependence on the sizes of interacting dipoles,  the generating
functional degenerates to the generating function that has the form
\beq \label{GAZ} Z_0(u|Y)\,\,=\,\,\sum_{n =
0}^{\infty}\,\,P_n(Y)\,u^n \eeq where  $P_n(Y)$ is the  probability
to find $n$-dipoles (or/and $n$-Pomerons) at rapidity $Y$. For   the
probabilities $P_n(Y)$ we can write Markov chain, namely
\cite{MUCD,L1,L2} \bea \label{MARCH} \dfrac{d P_n(Y)}{d Y}
\,\,&=&\,\,-\Gamma(1 \to 2)\,\,n\,\,P_n(Y)\,\,+\,\,\Gamma(1 \to
2)\,(n -1)\,P_{n -1}(Y) \, \eea \eq{MARCH} has a simple structure:
for the  process of dipole splitting we see two terms. The first one
with the negative sign  describes a decrease of probability $P_n$
due to the process of splitting  of dipoles. The second term with
positive sign is  responsible for the increase of the probability
due to the same processes of dipole interactions.

\eq{MARCH} can be re-written in the elegant form of the master
equation for $Z_0$, namely,

\beq \label{GA1}
\dfrac{\partial Z_0(u|Y)}{\partial Y}\,\,\,=-\Gamma(1 \to 2)\,\, u\,(  1 - u)\,\,\,
\dfrac{\partial
Z_0(u|Y)}{\partial
u}
\eeq
where $\Gamma(1 \to 2)\,\,\propto\,\,\bas$ in QCD.
The initial and boundary conditions look  as follows
\bea \label{IBC}
 \mbox{initial condition:} &~~ &Z_0(u | Y=0)\,\,=\,\,u\,; \nonumber \\
 \mbox{boundary  condition:}&~~ &Z_0(u=1 | Y)\,\,=\,\,1\,.
 \eea
 In \eq{IBC} the initial condition means that we are studying the evolution of one dipole, while the boundary condition
 follows from the normalization of the sum of probabilities.
 With this initial condition the linear differential equation \eq{GA1} can be written as non-linear one
\beq \label{Znonlin}
\frac{\partial Z_0(u|Y)}{\partial Y}\,\,\,=-\Gamma(1 \to 2)\,\, Z_0(u|Y)+\Gamma(1 \to 2)\,\, Z_0^2(u | Y)
\eeq

Introducing scattering amplitude \beq \label{Amp1}
N_0(\gamma|Y)=Im
A_{el}=-\sum_{n=1}^\infty \frac{(-1)^n}{n!}\frac{\partial^n
Z_0(u|Y)}{\partial u^n}|_{u=1}\gamma^n \eeq with $\gamma$ being a
scattering amplitude of interaction of single dipole with the
target, we can find \eq{GA1} for the amplitude \cite{KOVG}. \beq
\label{Amp2} \frac{\partial N_0(\gamma|Y)}{\partial Y}=\Gamma(1 \to
2)(\gamma-\gamma^2)\frac{\partial N_0(\gamma|Y)}{\partial \gamma}
\eeq

 Using initial condition $N_0(\gamma|Y=0)=\gamma$ one can rewrite \eq{Amp2} as non-linear equation for the amplitude
 \beq \label{Amp3}
 \frac{\partial N_0(\gamma|Y)}{\partial Y}=\Gamma(1 \to 2)N_0(\gamma|Y)-\Gamma(1 \to 2)N^2_0(\gamma|Y)
 \eeq
 \eq{Amp3} is easy to solve in this model and the solution has been found in   \cite{KL,BGLM,BORY}.
  It was noticed in   \cite{BK}   that one can get \eq{Amp3} directly from \eq{Znonlin} by substitution
  $N_0(\gamma|Y)=1-Z_0(1-\gamma | Y)$. Using this fact we can go back to \eq{GA1} and write it as
  \beq \label{Zgamma}
  \frac{\partial Z(1-\gamma | Y)}{\partial Y}\,\,\,=\Gamma(1 \to 2)\,\, \left(\gamma\frac{\partial
Z(1-\gamma | Y)}{\partial
\gamma}-  \gamma^2\frac{\partial
Z(1-\gamma | Y)}{\partial
\gamma} \right)
  \eeq


\subsection{The generating functional for the multiparticle production: definition and linear evolution equation}

  Here we would like to develop a method that will allow us not only to find the cross sections of the diffractive production but also to consider all processes  of multi-particle production at high energy. Having this goal in mind, we propose a generalization of the generating functional given by \eq{GAZ}, namely,  we introduce a new generating functional $Z\Lb u,v|Y \Rb$  as follows
 \beq \label{NZ}
Z\Lb u,v|Y \Rb=  \sum^{\infty}_{n=0,m=0} P\Lb n,m|Y \Rb\,u^n\,v^m
\eeq where $P\Lb n,m|Y \Rb$ is a probability to find $n$ uncut
Pomerons and $m$ cut Pomerons. Directly from \eq{NZ} and from the
fact that  $P\Lb n,m|Y \Rb$  is a probability we find the first
boundary condition
 \beq \label{ZNBC1} Z\Lb u=1,v=1|Y \Rb = 1\, \eeq

To find the second boundary condition we  can use the full form of
the $s$-channel unitarity constraint. Assuming that the scattering
amplitude is pure imaginary at high energy, this constraint looks as
follows \beq \label{FUC} 2\,N(s,b)\,\,=\,\,|N(s,b)|^2\,\,+\,\,G_{in}
\eeq where the first term in the l.h.s. is the elastic term with no
cut Pomerons,
 while the second is the total contribution of the inelastic processes
 (in other words, sum over all cut Pomerons).  Using functionals $Z_0( u | Y) $ and $Z\Lb u,v|Y \Rb$ we can
calculate the left and right hand sides of \eq{FUC}, namely,
\bea
N_0\,\,&=&\,\,1 \,\,-\,\,Z_0(1 - \gamma | Y)\,\,\,\, \mbox{ (see   \cite{BK})}\,\,; \label{ZVA1}\\
|N(s,b)|^2\,\,+\,\,G_{in}\,\,&=&\,\,2\,\Lb 1\,\,\,-\,\,Z\Lb 1\, -\, \gamma ,1\,-\,\gamma_{in} |Y \Rb\Rb \label{ZVA2}
\eea
where $\gamma  \,\,=\,\,N(s=0,b)$ is the imaginary part of the scattering amplitude of a dipole with the target at low energies, while $\gamma_{in} $ is the inelastic contribution ($|N(s,b)|^2\,\,+\,\,G_{in}$) for interaction of a dipole with the target at low energy. Generally speaking both these amplitudes are arbitrary and have to be calculated from  non-perturbative  QCD, however, assuming the low energies are not very low and we can apply the relation of \eq{UCP}
we see that $2\,\gamma\,\,=\,\,\gamma_{in}$.
 Using this relation we can re-write \eq{ZVA1} and \eq{ZVA2}  in the form

\beq \label{ZNBC2}
\,Z_0(u | Y)\,\,\,=\,\,\,Z\Lb u, u|Y \Rb
 \eeq

The initial condition depends on what we want to calculate. This is the main advantage of the generating functional that it allows us to calculate everything. For example, the  cross section of single diffraction   integrated over all produced masses
($\sigma_{sd}$)
 we can find just calculating $Z\Lb u, v=0|Y  - Y_0 \Rb$  for  the initial condition in the form
\beq \label{ZNIC1}
Z\Lb u, v|Y=Y_0 \Rb\,\,\,=\,\,\,v
\eeq
  The cross section is equal to
\beq \label{ZNIC2}
\sigma_{sd}\Lb \gamma(Y_0)|Y-Y_0) \Rb\,\,\,=\,\,\,1 \,\,-\,\,Z\Lb u=1 -\gamma, v=0|Y-Y_0 \Rb
\eeq

The main idea of this paper is to introduce cut Pomeron being split into three different states. By analogy with \eq{Zgamma} we can relate to each process correspondent term of differential equation
\bea
&&  \displaystyle{\not }P    \,\,\to \,\,  \displaystyle{\not }P  \,\,+\,\,  \displaystyle{\not }P  \hspace{0.5cm} \sim  \hspace{0.5cm}\gamma_{in}^2\,\frac{\partial Z}{\partial \gamma_{in}}\,\,; \label{DDD1}\\
&&   \displaystyle{\not }P \,\,\to \,\,  \displaystyle{\not }P \,\,+\,\, P \hspace{0.5cm} \sim  \hspace{0.5cm} \gamma\, \gamma_{in}\,\frac{\partial Z}{\partial \gamma_{in}}\,\,; \label{DDD2}\\
&&  \displaystyle{\not }P \,\,\to \,\,P \,\,+\,\,P \hspace{0.5cm} \sim  \hspace{0.5cm} \gamma^2\,\frac{\partial Z}{\partial \gamma_{in}}\,\,; \label{DDD3}
\eea
where the notation $ \displaystyle{\not }P $ and $P$ stand for cut and uncut Pomeron respectively and only second order terms in $\gamma$ ($\gamma_{in}$) are shown.
The next step it to use AGK cutting rules to write the resulting evolution equation
\beq \label{ZgammaIn}
\frac{\partial Z}{\partial Y}=\Gamma(1 \to 2)(\gamma-\gamma^2)\frac{\partial Z}{\partial \gamma}+
\Gamma(1 \to 2)(2\gamma^2-4\gamma\gamma_{in}+\gamma^2_{in}+\gamma_{in})\frac{\partial Z}{\partial \gamma_{in}}
\eeq
It can be easily seen that the second term reproduces the first term for $2\,\gamma=\gamma_{in}$.
We note that $u=1-\gamma$ and $v=1-\gamma_{in}$ and  thus \eq{ZgammaIn} for $u$ and $v$ reads as
\beq \label{Zuv}
\frac{\partial Z}{\partial Y}=-\Gamma(1 \to 2)u(1-u)\frac{\partial Z}{\partial u}-
\Gamma(1 \to 2)(2u^2-4uv+v^2+v)\frac{\partial Z}{\partial v}
\eeq

The description in terms of generating function becomes clear for
Markov chain. We use the definition of generating function given by
\eq{NZ} to find the differential equation for probabilities \bea
\frac{\partial P\Lb n,m|Y \Rb}{\partial(\,\Gamma(1 \to 2)\,\, Y)}\,\,\,&\,\,\,\,\,\,\,=\,\,\,\,\,\,& \nonumber\\
\Lb P\,\,\to \,\,P \,\,+\,\,P \,  \Rb & &- \,\,n\,P\Lb n,m|Y \Rb\,\,+\,\,(n - 1)\,P\Lb n-1,m|Y \Rb  \label{MCZN1} \\
\Lb \displaystyle{\not }P\,\,\to \,\,\displaystyle{\not }P \,\,+\,\,\displaystyle{\not }P \,  \Rb & & +\,\,m\,\,P\Lb n,m|Y \Rb \,\,-\,\,\,(m - 1)\,P\Lb n ,m - 1 |Y \Rb\,  \label{MCZN2}\\
\Lb \displaystyle{\not }P\,\,\to \,\,\displaystyle{\not }P \,\,+\,\,P\,  \Rb & & \,-\,\,4\,m\,\,P\Lb n,m|Y \Rb +\,\,4\,m\, P\Lb n-1,m|Y \Rb   \,\, \label{MCZN3}\\
\Lb \displaystyle{\not }P\,\,\to \,\,P\,\,+\,\,P \,  \Rb & &\,\,+\,\,2\,\,m\,\,P\Lb n,m|Y \Rb  \,\,-\,\,2\, ( m + 1)\,P\Lb n - 2 ,m + 1 |Y \Rb \,\, \label{MCZN4}
\eea
where each line describes specific process of Pomeron splitting discussed above.
 It is  instructive to compare this equation to \eq{MARCH}.  Each of  \eq{MCZN1} - \eq{MCZN4}
 consists of two terms: one  describes to increase of probability to find $n$-Pomerons due
  to decay of one Pomeron to two and one is responsible for the decrease
  of this probability since one of $n$ Pomerons can decay. In all equations, except \eq{MCZN2} and \eq{MCZN4},
  the increase  leads to the plus sign and a decrease to the minus sign.  However, in \eq{MCZN2}  and \eq{MCZN4}
   the signs are opposite in accordance with the AGK cutting rules (see \fig{agk3p}).
    In this case we have to say that decay $\displaystyle{\not }P\,\,\to \,\,P \,\,+\,\,P\,  $ and
$ \displaystyle{\not }P\,\,\to\,\, \displaystyle{\not
}P\,\,+\,\,\displaystyle{\not }P $  have  negative amplitudes. It
should be stressed that in terms of the amplitude (see
\eq{ZgammaIn}) we obtain
 positive cross sections with different multiplicity of produced particles.

 Equations \eq{MCZN1} -\eq{MCZN4} give clear probabilistic interpretation of all Pomeron splitting processes
 under discussion.

It turns out that \eq{Zuv}, being a typical Liouville equation, has the solutions that depends on
 two variables $\xi_1=\Gamma(1 \to 2)\,\, Y +\ln \frac{u}{1-u}$ and
$\xi_2=\Gamma(1 \to 2)\,\, Y +\ln \frac{2u-v}{1-(2u-v)}$. The
general solution of this equation is given by an arbitrary function
of variables $\xi_1$ and $\xi_2$, which in our case can be written
as sum of two functions, namely, \beq \label{F1F2} Z=F_1
\left\{\Gamma(1 \to 2)\,\, Y +\ln
\frac{u}{1-u}\right\}+F_2\left\{\Gamma(1 \to 2)\,\, Y +\ln
\frac{2u-v}{1-(2u-v)}\right\} \eeq For our initial condition
\eq{ZNIC1} the solution reads \beq \label{SolZ} Z=\dfrac{2u
e^{-\Gamma(1 \to 2)  Y}}{1+u(e^{-\Gamma(1 \to 2)
Y}-1)}-\dfrac{(2u-v) e^{-\Gamma(1 \to 2)  Y}}{1+(2u-v)(e^{-\Gamma(1
\to 2)\,\, Y}-1)} \eeq

One can easily see that this solution satisfies both initial \eq{ZNIC1} and boundary \eq{ZNBC1} conditions.

\subsection{The generating functional for the multiparticle production: non-linear equation}
Using our initial condition \eq{ZNIC1} we can rewrite linear differential equation \eq{Zuv} as non-linear one.
We use the fact mentioned above, namely, that \eq{Zuv} is differential equation of two variable $\xi_1$ and $\xi_2$ and, thus, has no separate dependence on $Y$. This means that the differential equation \eq{Zuv} written at some rapidity $Y$ keeps the same form for any rapidity. We pick initial rapidity $Y=0$ and substitute generating function given by \eq{ZNIC1} into \eq{Zuv}
\beq \label{NonLin1}
\frac{\partial Z}{\partial Y}=0-
\Gamma(1 \to 2)(2u^2-4uv+v^2+v)
\eeq
Now use initial condition  from \eq{IBC} for generating function with no cut Pomerons. In terms of $Z_0$ and $Z$ \eq{NonLin1} reads
\beq \label{NonLin2}
 \frac{\partial Z}{\partial Y}=-
\Gamma(1 \to 2)(2Z_0^2-4Z_0Z+Z^2+Z)
\eeq
We identify scattering amplitude and diffractive cross section with $N_0(\gamma|Y)=1-Z_0(1-\gamma | Y)$ and
 $N(\gamma,\gamma_{in}|Y)=1-Z(1-\gamma,1-\gamma_{in}|Y)$, respectively. For $N_0(\gamma| Y)$ and $N(\gamma,\gamma_{in}|Y)$ \eq{NonLin2} reads
\beq \label{NonLin3}
\frac{\partial N(\gamma,\gamma_{in}| Y)}{\partial Y}=-
\Gamma(1 \to 2)\left\{2N^2_0(\gamma|Y)-4N_0(\gamma|Y)N(\gamma,\gamma_{in}|Y)+N^2(\gamma,\gamma_{in}|Y) +N(\gamma,\gamma_{in}|Y)\right\}
\eeq
This equation has the same form as the equation for the diffraction production\footnote{ The difference in overall  minus sign corresponds to different definitions of rapidity variable moving in opposite direction.} obtained by Kovchegov and Levin  (see \eq{KLND}),  but it is written for a general functional and describes not only diffractive production but also the processes of particle production  with any  value of multiplicity. The fact that \eq{KLND} has a simple generalization on the general case of QCD
gives us a hope to generalize this equation.

\subsection{The generating functional for the multiparticle production: consistency with the AGK cutting rules}
 We want to check the consistency of our solution \eq{SolZ} with the AGK cutting  rules in an explicit way. To do this we calculate cross section of a process $\sigma^{(k)}$ with a given multiplicity $k$ from both generating function given by \eq{SolZ} and  directly from the AGK cutting rules, and compare them. We define cross section  with multiplicity $k$ as
 \beq \label{1Mult}
\sigma^{(k)}=\frac{1}{k!}\frac{\partial^k N(\gamma,\gamma_{in}|Y)}{\partial \gamma_{in}^k}\mid_{\gamma_{in}=0}\cdot\gamma^k_{in}
 \eeq
where $N(\gamma,\gamma_{in}|Y)=1-Z(1-\gamma,1-\gamma_{in}|Y)$.

As an example we pick multiplicity to be that of one cut Pomeron, i.e. $k=1$ for any number of uncut Pomerons.
From \eq{1Mult} with $k=1$ we get
\beq \label{ZAGK}
\sigma^{(1)}=\frac{\gamma_{in}e^{\Gamma(1 \to 2)  Y}}{(1+2\gamma(e^{\Gamma(1 \to 2)  Y}-1))^2}
\eeq
On the other hand we can use coefficients for multiple Pomeron exchange from the AGK cutting rules \cite{AGK}. In this case cross section for multiplicity of  $k$ cut Pomerons reads
\beq \label{AGKAGK}
\sigma^{(k)}=\sum_{n=0}^\infty (-1)^{n-k}C^n_k(2\gamma)^n e^{n\Gamma(1 \to 2)  Y}
\eeq
where $e^{\Gamma(1 \to 2)  Y}$ stands for Pomeron propagator.

For $k=1$ we sum over $n$ in \eq{AGKAGK} and obtain
\beq \label{AGKAGK2}
\sigma^{(1)}=\frac{2\gamma e^{\Gamma(1 \to 2)  Y}}{(1+2\gamma e^{\Gamma(1 \to 2)  Y})^2}
\eeq

In high energy limit $\gamma_{in}=2\gamma$ and $e^{\Gamma(1 \to 2)  Y}-1\simeq  e^{\Gamma(1 \to 2)  Y}$, thus
\eq{ZAGK} reproduces \eq{AGKAGK2}. It can be easily shown that this holds for any value of $k$.

We can further compare cross sections obtained from generating
function and direct summation of fan diagrams using  AGK rules
\cite{Boreskov:2005ee}.
In our approach this corresponds to \beq \label{sdGenFun}
\sigma_{sd}=N(\gamma,\gamma_{in}=0|Y)  \hspace{2cm}
\sigma_{in}=N(\gamma,\gamma_{in}=2\gamma_{in}|Y) \; \eeq

where $N(\gamma,\gamma_{in}|Y)=1-Z(1-\gamma,1-\gamma_{in}|Y)$. The
resulting expressions are identical and given by

\bea
  \sigma_{sd}=\frac{2\gamma^2 e^{\Gamma(1 \to 2)  Y}(e^{\Gamma(1 \to 2)  Y}-1)}{(1+\gamma( e^{\Gamma(1 \to 2)  Y}-1))
 (1+2\gamma (e^{\Gamma(1 \to 2)  Y}-1))} \nonumber
 \eea
 \bea
\sigma_{el}=\frac{2\gamma e^{\Gamma(1 \to 2)  Y}}{1+2\gamma
(e^{\Gamma(1 \to 2) Y}-1)}
 \eea
\bea
 \sigma_{tot}=\frac{2\gamma e^{\Gamma(1 \to 2)  Y}}{1+\gamma (e^{\Gamma(1 \to 2)  Y}-1)} \nonumber \;\; .
 \eea

\section{Pomeron loops}\label{sec:Loops}
\subsection{Evolution equation with loops}
Now we want to account for contributions of Pomeron loops. This
problem has already been solved for a case with no cut Pomerons
\cite{L2}. The master equation for no cut Pomerons is given by \beq
\label{MastU} \frac{\partial Z}{\partial Y}=-\Gamma(1 \to
2)u(1-u)\frac{\partial Z}{\partial u}+ \frac{1}{2}\Gamma(2 \to
1)(u-u^2)\frac{\partial^2 Z}{\partial u^2} \eeq Unfortunately, one
cannot  generalize \eq{Zuv} for diffractive processes  by just
adding second order derivative terms by analogy with \eq{MastU}. The
reason for that is because this type of equation would include
diagrams that does not exist. For example, diagrams of the type
$\displaystyle{\not }P \rightarrow P\; +\;P \rightarrow P$ where cut
Pomeron splits to two uncut Pomerons with further merging to uncut
Pomeron are forbidden. To resolve this problem we introduce two
separate variables $w$ and $\bar{w}$ for uncut Pomeron being in
amplitude or conjugate amplitude, respectively. Naturally, these two
subsets of Pomerons evolve separately till the cut Pomeron is
introduced, alternatively, their evolutions mix only via cut
Pomeron. Using our previous discussions we can readily write this
new type of evolution equation based on \eq{Zuv} \bea \label{Zwv}
\frac{\partial Z}{\partial Y}= &-&\Gamma(1 \to
2)\left\{w(1-w)\frac{\partial Z}{\partial w}
- \bar{w}(1-\bar{w})\frac{\partial Z}{\partial \bar{w}} \right\} \nonumber \\
&-&
\Gamma(1 \to 2)(2w\bar{w}-2wv-2\bar{w}v+v^2+v)\frac{\partial Z}{\partial v}\\
&+&
\frac{1}{2}\Gamma(2 \to 1)\left\{(w-w^2)\frac{\partial^2 Z}{\partial w^2}
+(\bar{w}-\bar{w}^2)\frac{\partial^2 Z}{\partial \bar{w}^2} \right\} \nonumber \\
&-&\frac{1}{2}\Gamma(2 \to 1)\left\{ 2(v-w\bar{w}) \frac{\partial^2
}{\partial w \partial \bar{w}} +2(v-wv) \frac{\partial^2 }{\partial w
\partial v} +2(v-\bar{w}v) \frac{\partial^2 }{\partial \bar{w}
\partial v} +(v-v^2)\frac{\partial^2}{\partial v^2} \right\} Z
\nonumber \eea where the first and second lines are generalization
of \eq{Zuv} for variables $w$ and $\bar{w}$, the third line
corresponds to \eq{MastU} and the last line a little bit more
explanation. We have to introduce a new generating function \beq
\label{genw} Z(w, \bar{w},v|Y)=\sum_{k=0}\sum_{l=0}
\sum_{m=0}P(k,l,m|Y) w^k \bar{w}^l v^m \eeq where $P(k,l,m|Y)$ stands
for probability to find $k$ uncut Pomerons in the amplitude, $l$
uncut Pomerons in the conjugate amplitude and $m$ cut Pomerons at some rapidity $Y$. For
the last term of \eq{Zwv} we write Markov chain in a similar manner
we did it for \eq{MCZN1}-\eq{MCZN4}, namely, \bea \label{MarkovW}
\frac{\partial P\Lb k,l,m|Y \Rb}{\partial Y}\,\,\, &=&\\
\Lb P \, \to  \,P \,+ \,P \,  \Rb  &\,\,\,\,\,\,\,+\,\,\,\,\,\,\,&
\Gamma(1 \to 2)\left\{(k-1)P(k-1,l,m|Y) -k P(k,l,m|Y) \right\} \;\;     \nonumber \\
 \Lb \bar{P} \, \to  \,\bar{P} \,+ \,\bar{P} \,  \Rb &+&\Gamma(1 \to 2)\left\{(l-1) P(k,l-1,m|Y) -l P(k,l,m|Y) \right\} \;\;   \nonumber \\
\Lb \displaystyle{\not }P \, \to  \,\displaystyle{\not }P \,+ \,\displaystyle{\not }P\Rb
&-&\Gamma(1 \to 2)\left\{(m-1) P(k,l,m-1|Y) -m P(k,l,m|Y) \right\} \;\; \nonumber\\
\Lb \displaystyle{\not }P \, \to  \, P \,+ \,\displaystyle{\not }P \,  \Rb
&+&2\;\Gamma(1 \to 2)\left\{m P(k-1,l,m|Y) -m P(k,l,m|Y) \right\} \;\;\nonumber \\
\Lb \displaystyle{\not }P \, \to  \, \bar{P} \,+ \,\displaystyle{\not }P \,  \Rb
 &+& 2\;\Gamma(1 \to 2)\left\{m P(k,l-1,m|Y) -m P(k,l,m|Y) \right\} \;\;\nonumber
\\
\Lb \displaystyle{\not }P \, \to  \,P  \,+  \bar{P} \,  \Rb&-&2\;\Gamma(1 \to 2)\left\{(m+1) P(k-1,l-1,m+1|Y) -m P(k,l,m|Y) \right\} \;\;  \nonumber \\
\Lb  P \,+ \, P \to  \,P \,  \Rb &+& \frac{1}{2}\;\Gamma(2 \to 1)\left\{k(k+1) P(k+1,l,m|Y) -k (k-1) P(k,l,m|Y) \right\} \;\;  \nonumber \\
\Lb  \bar{P} \,+ \, \bar{P} \to  \,\bar{P} \,  \Rb &+& \frac{1}{2}\;\Gamma(2 \to 1)\left\{l(l+1) P(k,l+1,m|Y) -l (l-1) P(k,l,m|Y) \right\} \;\;  \nonumber \\
\Lb  \displaystyle{\not }P \,+ \, \displaystyle{\not }P \to  \,\displaystyle{\not }P \,  \Rb
&-&\frac{1}{2}\;\Gamma(2 \to 1)\left\{m(m+1) P(k,l,m+1|Y) -m (m-1) P(k,l,m|Y) \right\} \;\;
\nonumber \\
\Lb  \displaystyle{\not }P \,+ \, P \to  \,\displaystyle{\not }P \,  \Rb
&-&\frac{2}{2}\;\Gamma(2 \to 1)\left\{(k+1)m P(k+1,l,m|Y) -k m  P(k,l,m|Y) \right\} \;\;
 \nonumber \\
\Lb  \displaystyle{\not }P \,+ \, \bar{P} \to  \,\displaystyle{\not }P \,  \Rb
&-&\frac{2}{2}\;\Gamma(2 \to 1)\left\{(l+1)m P(k,l+1,m|Y) -l m  P(k,l,m|Y) \right\} \;\;\nonumber \\
\Lb  P \,+ \, \bar{P} \to  \,\displaystyle{\not }P \,  \Rb
&-&\frac{2}{2}\;\Gamma(2 \to 1)\left\{(k+1)(l+1) P(k+1,l+1,m|Y) -k l  P(k,l,m|Y) \right\} \;\;  \nonumber
\eea
where $\bar{P}$ denotes Pomeron in conjugate amplitude and  factor of $\frac{1}{2}$ accounts for a fact that the Pomerons are identical in this approach. Each line in \eq{MarkovW}
as in \eq{MCZN1}-\eq{MCZN4} has a clear probabilistic interpretation.
We multiply  \eq{MarkovW} by $w^k \bar{w}^l v^m$, sum over all $k$, $l$ and $m$, and using definition \eq{genw}
obtain \eq{Zwv}.

 Even in our simple model with no coordinate dependence \eq{Zwv} is too much complicated and its solution is still to be found.
 But before the solution is found we can see that putting $w=\bar{w}=u$, i.e. making no difference between uncut Pomerons,
 first two lines of \eq{Zwv} correctly reproduce \eq{Zuv}. Moreover, we can check it further and  using initial condition $Z(w, \bar{w},v|Y=0)=v$
 we can perform iterations
 \bea \label{iter1}
 \frac{\partial Z_1}{\partial Y}=-
\Gamma(1 \to 2)(2w\bar{w}-2wv-2\bar{w}v+v^2+v)
  \eea
  giving
  \bea \label{iter2}
 Z_1 =-
\Gamma(1 \to 2)(2w\bar{w}-2wv-2\bar{w}v+v^2+v)Y
  \eea
  At the next step

  \bea \label{iter3}
  Z_2 =
&+&\Gamma^2(1 \to 2)\left\{w(1-w)(2\bar{w}-2v)
- \bar{w}(1-\bar{w})(2 w -2v) \right\}\frac{Y^2}{2} \nonumber \\
&+&
\Gamma^2(1 \to 2)(2w\bar{w}-2wv-2\bar{w}v+v^2+v)(-2(w+\bar{w})+2v+1)\frac{Y^2}{2}
\\
&-&\frac{1}{2}\Gamma(1\to 2)\Gamma(2 \to 1)\left\{
2(v-w\bar{w})(-2)
+2(v-wv) 2
+2(v-\bar{w}v) 2
+(v-v^2)(-2)
\right\} \frac{Y^2}{2}  \nonumber
  \eea
 It is easy to see from \eq{iter3} that \eq{Zwv} correctly reproduces the sign and combinatorics coefficient of Pomeron loops
 in accordance with AGK cutting rules. For example, in the first term in the third line
 $-\frac{1}{2}\Gamma(1\to 2)\Gamma(2 \to 1)
2(v-w\bar{w})(-2)$, the term proportional to $v$ describes the Pomeron loop of the type
 $\displaystyle{\not }P \; \to P \,+ \, \bar{P} \to  \, \displaystyle{\not }P $ . This loop has factor of $2$ and brings
 plus sign as expected. Similarly, the terms proportional to $v$ in the second and third term of the last line correspond
 to loops of the type $\displaystyle{\not }P \; \to \displaystyle{\not } P \,+ \,  P \to  \, \displaystyle{\not }P $
 and $\displaystyle{\not }P \; \to \displaystyle{\not } P \,+ \, \bar{P} \to  \, \displaystyle{\not }P $, respectively.
 Each of them brings factor of $-2$, putting $P=\bar{P}$ we have a factor of $-4$ which is in a agreement with AGK cutting rules.
 Thus, we expect \eq{Zwv} properly include Pomeron loops  into evolution.

 At the end of this part we would like to mention that
  the problem has been already treated  by Ciafaloni  and Marchesini \cite{Ciafaloni:1975ph} many years ago,
  but in terms of RFT  Lagrangian. They separated uncut Pomerons in amplitude and conjugate amplitude by introducing
  different fields  ($\phi_+$ and $\phi_-$). It is interesting to notice that, just like in our case, the authors
   also have partial diagonalization of Lagrangian in terms of new variable $\phi_+ + \phi_- - i\phi_c$.
   By partial diagonalization one should understand diagonal vertices only in one rapidity direction.
   In our case this can be formulated as separation of evolutions of Pomerons corresponding to $w$, $\bar{w}$ and
   $w+\bar{w}-v$ in MFA.

\subsection{A new method of summation of the Pomeron loops: improved Mueller-Patel-Salam-Iancu approach}
Solution to \eq{Zwv} will give the generating functional. However, \eq{Zwv} is based on the probabilistic interpretation
  of the BFKL Pomeron calculus in terms of  Markov process. In this subsection we would like to suggest a different interpretation with a different procedure of summation of the Pomeron diagrams.

\FIGURE[ht]{
\centerline{\epsfig{file=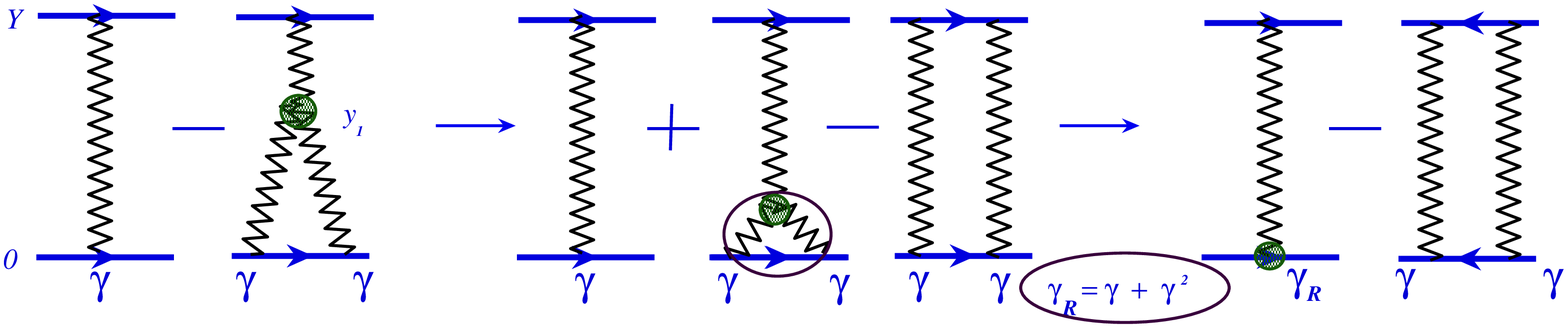,width=180mm,height=40mm}}
\caption{ The renormalization procedure  in the case of the simplest `fan' diagram. }
\label{dtrre}
}

Firstly we consider the simplest `fan' diagram of \fig{dtrre}. It can be calculated in an obvious way, namely,
\bea \label{MPSI1}
A\Lb\fig{dtrre}\Rb\,\,\,&=&\,\,\gamma\,G(Y-0)\,\,-\,\,\Delta\,\gamma^2\,\int^Y_0\,d\,y_1\,\,G(Y - y_1)\,G^2(y_1 - 0) \,\,\,\\
 &=&\,\,\gamma\,e^{\Delta Y}\,\,\,-\,\,\,\Delta\,\gamma^2\,\int^Y_0\,d\,y_1\,\,e^{\Delta \Lb Y + y_1  \Rb}\,\\
 &=&\,\,\gamma\,e^{\Delta Y}\,\,\,-\,\,\,\Delta\,\gamma^2\,\Lb \frac{1}{\Delta}\,e^{2 \,\Delta\,Y}\,\,\,-\,\, \frac{1}{\Delta}\,e^{\Delta\,Y}\Rb\,\,\,\nonumber \\
& =&\,\,-\,\,\gamma^2\,\,e^{2 \,\Delta\,Y}\,\,\,+\,\,\,(\gamma\,+\,\gamma^2)\,\,e^{ \Delta\,Y}
\,\,=\,\,-\,\,\gamma^2\,\,e^{2 \,\Delta\,Y}\,\,\,+\,\,\gamma_{R}\,\,e^{ \Delta\,Y}
 \nonumber
 \eea
where $\Gamma(1 \to 2)$ (see \eq{GA1})  is denoted as $\Delta$ , $\gamma$ is the amplitude of the Pomeron interaction
with the target and $G(Y - y)$ stands for the Green function of the Pomeron $G(Y - y)\,=\,\exp\Lb \Delta\,( Y - y)\Rb$.

As one can see, the integration over $y_1$ reduces the diagram in \fig{dtrre} to two contributions: the exchange of two non-interacting Pomerons and the exchange of one Pomeron with the renormalized vertex: $\gamma_R\,\,=\,\,\gamma\,\,+\,\,\gamma^2$. In \fig{dtrre1} is shown the Pomeron `fan' diagram of the second order which we have to integrate over two rapidities $y_1$ and $ y_2$.  The result is
\par
$
A\Lb\fig{dtrre1}\Rb\,\,\,=
$
\bea
\,\,\,\,\,&=&2\,\Delta^2\,\gamma^3\,\int^Y_0\,d\,y_1\,\,\int^{y_1}_0\,d\,y_2\,G(Y - y_1)\,G(y_1 - 0)\,G(y_2 -0)\,\,G^2(y_2 - 0) \,\, \label{MPSI2}\\
 &=&\,\,\,2\,\Delta^2\,\gamma^3\,\int^Y_0\,d\,y_1\,\int^{y_1}_0\,d\,y_2\,\,e^{\Delta \Lb Y + y_1  + y_2 \Rb}\,\,
 \,=\,\,\,2\,\Delta^2\,\gamma^3\,\Lb \frac{1}{2\,\Delta^2}\,e^{3 \,\Delta\,Y}\,\,\,-\,\, \frac{1}{\Delta^2}\,e^{2\,\Delta\,Y}\,\,+\,\, \frac{1}{2\,\Delta^2}\,e^{\,\Delta\,Y}\Rb\,\,\, \nonumber
 \eea
\FIGURE[ht]{
\centerline{\epsfig{file=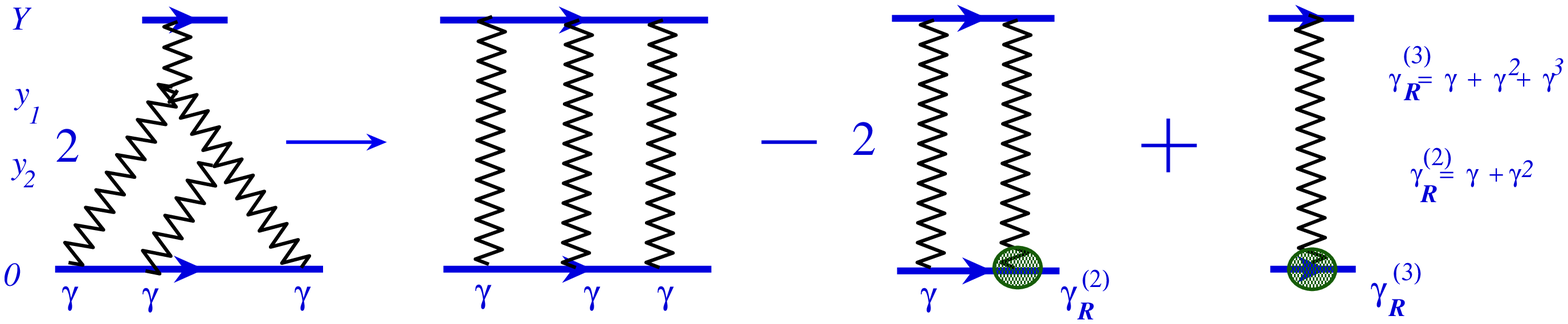,width=180mm,height=40mm}}
\caption{ The renormalization procedure  in the case of the  `fan' diagram of the second order. }
\label{dtrre1}
}
Adding the contributions of this diagram and the diagrams of \fig{dtrre} we obtain
\bea \label{MPSI3}
&&\,A\Lb\fig{dtrre}\Rb\,\,\,+\,\,\,A\Lb\fig{dtrre1}\Rb\,\,\,\,= \\
&& =\,\,-\,\,\gamma^3\,\,e^{3 \,\Delta\,Y}\,\,\,-\,\,\,\gamma\,(\gamma\,+\,\gamma^2)\,\,e^{2 \Delta\,Y}
\,\,+\,\,(\gamma \,\,+\,\,\gamma^2\,\,+\,\,\gamma^3)\,e^{3 \Delta\,Y} \,
\,\,=\,\,\,\,\gamma^3\,\,e^{3 \,\Delta\,Y}\,\,\,-\,\,2\,\gamma\gamma^{(2)}_{R}\,\,e^{2 \Delta\,Y}\,\,+\,\,\gamma_R\,e^{\Delta\,Y} \nonumber
 \eea
Therefore, one can see that the scattering amplitude can be rewritten as exchanges of the Pomerons without interaction between them but with  Pomeron - particle vertex. In the dipole model this vertex has a meaning of the amplitude of two dipole interaction in the Born approximation of perturbative QCD.

These two examples illustrates our main idea: the BFKL Pomeron
calculus in zero transverse dimensions can be viewed as the theory
of free, non-interacting Pomerons whose interaction with the target
has to be renormalized. It is easy to see that in the MFA we can
rewrite the master equation (see \eq{GA1} and \eq{Amp2}) in the form
\beq \label{MPSI4} \frac{\partial \,N_0\Lb
\gamma_R|Y\Rb}{\partial\,Y}\,\,\,\,=\,\,\Gamma(1 \to
2)\,\gamma_R\,\frac{\partial \,N_0\Lb
\gamma_R|Y\Rb}{\partial\,\gamma_R} \eeq with \beq \label{MPSI5}
\gamma_R\,\,\,=\,\,\frac{\gamma}{1 \,\,-\,\,\gamma} \eeq

\FIGURE[ht]{\begin{minipage}{65mm}{
\centerline{\epsfig{file=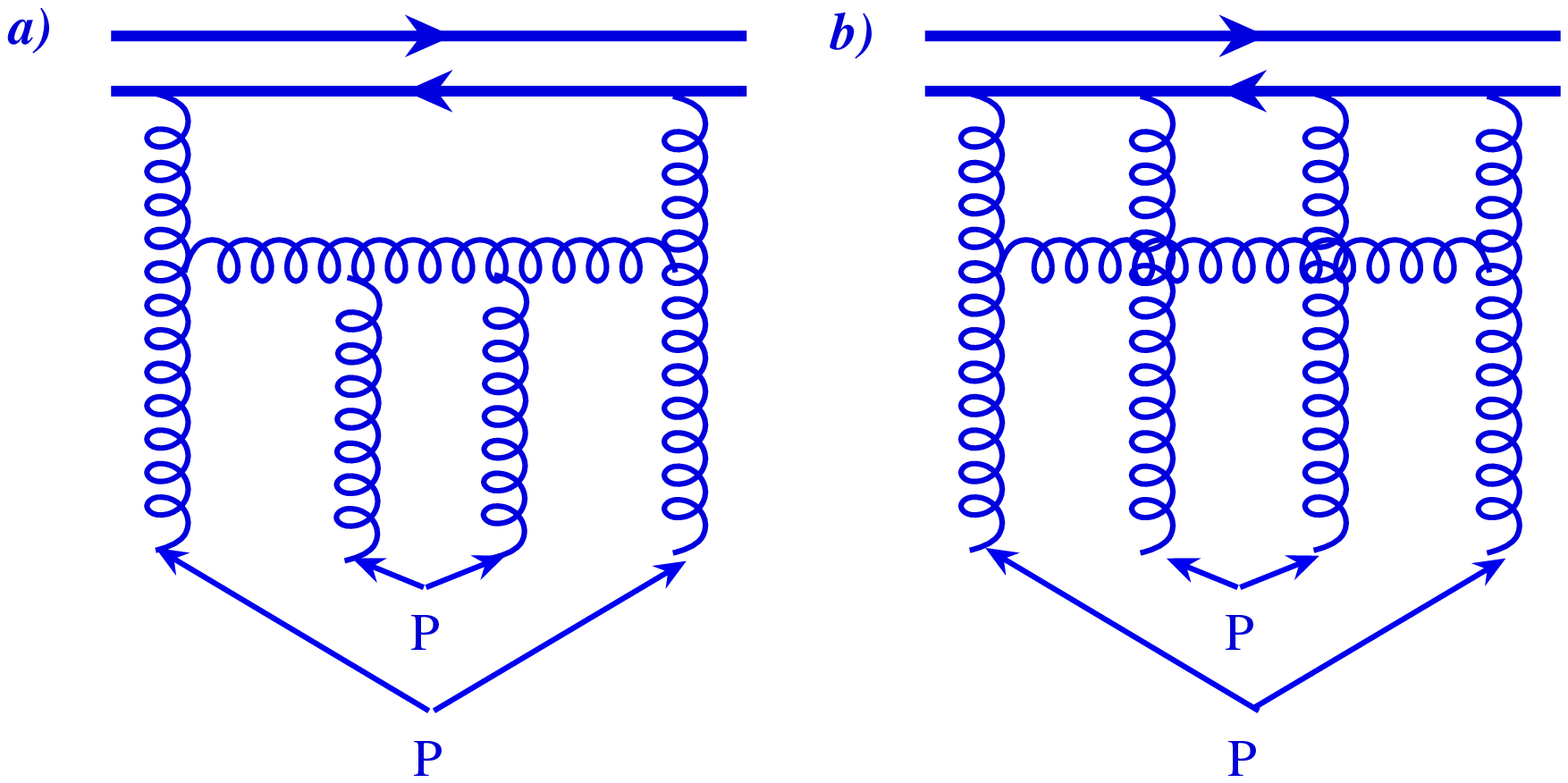,width=60mm}}
}
\end{minipage}
\caption{The diagram that illustrates the initial condition of \protect\eq{MPSIIC}.}
\label{diic}
}

 The way \eq{MPSI5} has started to build in perturbation expansion we have shown in \eq{MPSI1} and \eq{MPSI2}.

The general solution of \eq{MPSI4} is the system of non-interacting Pomerons and the scattering amplitude can be found in the form
\beq\label{MPSI6}
N_0\Lb \gamma_R|Y\Rb\,\,\,=\,\,\,\,\sum_{n=1}^{\infty}\,\,\,\,\,C_n\,\,\gamma^n_R\,\,G^n(Y - 0)
\eeq
where coefficients $C_n$ could be found from the initial conditions, namely, from the expression for the low energy amplitude. In particular, the initial condition
\beq \label{MPSIIC}
 N_0\Lb \gamma_R|Y=0\Rb\,\,=\,\,\gamma\,\,=\,\,\gamma_R/(1 + \gamma_R)
\eeq
  generates $C_n\,\,=\,\,(-1)^n$ and the solution is
\beq \label{MPSI7}
N_0\Lb \gamma_R|Y\Rb\,\,\,=\,\,\,\frac{\gamma_R\,e^{\Delta\,Y}}{1\,\,\,\,+\,\,\,\gamma_R\,e^{\Delta\,Y}}
\eeq
The initial condition of \eq{MPSIIC} has a very simple physics behind it. It describes the independent (non-correlated)  production of the Pomerons at low energy but with the only condition that one Pomeron lives shorter time than the second one (see \fig{diic}).  If all $n$ Pomerons were emitted by the same dipole (see \fig{diic}-b)
this condition leads to Glauber  factor $1/n!$ leading to $N_0 (\gamma_R|Y=0) =  \gamma\,=\,1 - \,\exp( - \gamma_R)$.
instead of \eq{MPSIIC}.
 However, if Pomerons are produced as the consequent decays (see
\fig{diic}-a) the factor is equal to 1. $(-1)^n$ comes from the Glauber screening, resulting in \eq{MPSIIC}. In QCD we have strong evidence
that the second case is correct \cite{BART}.

The analysis of enhanced diagrams we start from the first diagram of \fig{denre}. It leads to the following contribution
\bea \label{MPSI8}
&&A\Lb \fig{denre}\Rb\,
    \,\,= \\
&& \,\,-\,\Delta^2\,\gamma^2\,\,\int^Y_0\,d\,y_1\,\int^{y_1}_0\,d\,y_2\,G(Y - y_1)\,G^2(y_1 - y_2)\,G(y_2 - 0) \,\,\,\nonumber \\
&&=\,\,\,-\,\Delta^2\,\gamma^2\,\,\int^Y_0\,d\,y_1\,\int^{y_1}_0\,d\,y_2\,\,e^{\Delta\,(Y + y_1 - y_2)}\nonumber\\     &&=\,\,-\,\,\,\gamma^2\,e^{2\,\Delta\,Y}\,\,+\,\,\gamma^2\,e^{\,\Delta\,Y}\,\,+\,\,\Delta\,\gamma^2\,Y\,
      e^{\,\Delta\,Y} \nonumber
      \eea
      where $\Gamma(2 \to 1) \,\,=\,\,\Delta\,\gamma^2$ (see  \cite{IT}).

Adding \eq{MPSI8} to the exchange of the one Pomeron we obtain that  the exchange of one Pomeron and the enhanced diagram of \fig{denre} can be written in close form
\beq \label{MPSI9}
\mbox{One Pomeron exchange} \,+\,A\Lb \fig{denre}\Rb\,\,\,=\,\,\,\gamma_R\,e^{\,\Delta_R\,Y} \,\,-\,\,\gamma^2\,e^{2\,\Delta\,Y}
\eeq
with
\beq \label{MPSI10}
\gamma_R\,\,=\,\,\gamma^{(2)}\,\,=\,\,\gamma\,\,+\,\,\gamma^2\,;\,\,\,\,\,\,\,\,
\Delta_R\,\,=\,\,\Delta\,\,+\,\,\gamma\,\Delta\,;
\eeq
It is easy to see the \eq{MPSI8} can be viewed as the expansion to  first order of \eq{MPSI9}

Therefore, the Pomeron  loops  can be either large (of  the order of $Y$) and they can be considered   as un-enhanced diagrams ; or  small (of the order of $1/\Delta$) and they can be treated as the renormalization of the Pomeron intercept.

In QCD\,\,$\Delta \,\propto \,\bas$ while $\gamma\,\,\propto\,\as^2$. Therefore, the renormalization of the Pomeron intercept $\Delta$ is proportional to $\as^3$.   We can neglect this contribution since  (i)  there a lot of $\as^2$ corrections to the kernel of the BFKL equation that  are much larger than this contribution; and (ii) in the region of $Y\,\,\ll\,\,1/\as^2$ , where we can trust our Pomeron calculus (see introduction)  $(\Delta_R\,-\,\Delta)\,Y\,\ll \,\,1$.

Concluding this analysis we can claim that the BFKL Pomeron calculus
in zero dimensions is a theory of non-interacting Pomerons with
renormalized vertices of Pomeron-particle interaction.  In the
dipole language, it means that we have a system non-interacting
Pomerons with a specific hypothesis on the amplitude of the dipole
interactions at low energy. For the problem that we are solving
here, namely, when we have one bare Pomeron at low energy, this
amplitude is determined by \eq{MPSIIC}.

For such a system we can calculate the scattering amplitude using a
method suggested by Mueller, Patel, Salam and Iancu and developed in
a number of papers (see   \cite{MPSI,KOLE,BOR,L4,KOVG,KLTM,KKLM} and
references therein). This method suggests that the scattering
amplitude can be calculated using the $t$-channel unitarity
constraints which is written in the following way  (assuming that
amplitudes at high energy  are pure imaginary, $N\,=\,Im\,A$): \beq
\label{MPSIUN} N([\dots]|Y)\,\,=\,\,N([\dots]|Y - Y';P \to
nP)\,\bigotimes \,N([\dots]|Y';P \to nP) \eeq where $\bigotimes$
stands for all needed integrations while   $[\dots]$ describes all
quantum numbers (dipole sizes and so on ).

The correct implementation of this leads in  our case  to the
following formula (see also   \cite{KOVG,KOLE,KKLM}) \beq
\label{MPSIEQ} N^{MPSI}_0\Lb Y \Rb\,\,\,=\,\,\,1\,\,\,-\,\,\exp
\left\{\,-\,\gamma^{BA}\,\frac{\partial}{\partial
\gamma^1_{R}}\,\frac{\partial}{\partial
\gamma^2_R}\,\right\}\,N^{MFA}\Lb \gamma^1_R|Y - Y' \Rb\, N^{MFA}\Lb
\gamma^2_R|Y'\Rb|_{\gamma^1_R\,=\,\gamma^2_R \,=\,0} \eeq where
$N^{MFA}\Lb  Y,\gamma_R \Rb$ is given by \eq{MPSI7}(see also
\eq{MPSI4})   in the mean field approximation and
$\gamma^{BA}\,\propto\,\as^2$ is the scattering amplitude at  low
energies which is described by the Born approximation in
perturbative QCD. The difference of \eq{MPSIEQ} from the original
MPSI approach is the fact that this equation does not depend on the
value of $Y'$  and, because of this, we do not need to choose $Y' =
Y/2$ for the best accuracy.

Substituting \eq{MPSI7} in \eq{MPSIEQ} we obtain \bea
\label{MPSIEQ1} N^{MPSI}_0\Lb \gamma^{BA}|Y\Rb\,\,\,
&=&\,\,1\,\,-\,\,\exp \Lb \frac{1}{\gamma^{BA} e^{ \Gamma(1 \to 2) Y
}}\Rb\,\frac{1}{\gamma^{BA} e^{ \Gamma( 1 \to 2)Y }}\,\, \Gamma\Lb
0,\frac{1}{\gamma^{BA} e^{ \Gamma(1 \to 2) Y } }\Rb \eea $\Gamma \Lb
0,x \Rb$ is the incomplete gamma function  (see  formulae {\bf 8.350
- 8.359} in   \cite{RY}).

We claim that \eq{MPSIEQ1} is the solution to our problem. One can
easily see that $N_0\Lb \gamma|Y\Rb\,\to\,1$ at high energies in
contrast to the exact solution with Hamiltonian of \eq{I1}. The
exact solution leads to the amplitude that vanishes at high energy
(see  \cite{AMCP,BMMSX}). As have been mentioned the solution
depends crucially on the initial condition for the scattering
amplitude at low energies. For \eq{MPSIEQ1} this amplitude is equal
to \beq \label{MPSIEQ2} N^{MPSI}_0\Lb
\gamma|Y=0\Rb\,\,\,=\,\,\sum^{\infty}_{n=1}\,(-1)^{n+1}\,\,n!\,\Lb\gamma^{BA}
\Rb^n \eeq with $\gamma^{BA}\,\propto\,\as^2$. This equation reminds
us the ultraviolet renormalons contribution and calls for better
understanding of the non-perturbative contribution.

 We can  rewrite \eq{MPSIEQ} in more convenient form using the Cauchy formula for the derivatives, namely,
\bea
\frac{\partial^n Z^{MFA}(\gamma_R|Y)}{\partial \gamma^n_R}\,\,&=&\,\,\,n!\,\frac{1}{2\,\pi i}\oint_C\,\,\frac{Z^{MFA}(\gamma'_R| Y)}{\gamma'^{n+1}_R}\,d\,\gamma'_R;\label{PL2}
\eea
Contour $C$ in \eq{PL2} is a circle with a small radius around $\gamma_R = 0$. However, since function $Z$ does not  grow at large $\gamma_R$ for $n \leq 1$ we can close our contour $C$ on the singularities of function $Z$. We will call this new contour $C_R$.
\bea
&&N^{MPSI}_0\Lb  Y \Rb\,=\,1\,\,\,-\,\,\exp \left\{\,-\,\gamma^{BA}\,\frac{\partial}{\partial \gamma^1_{R}}\,\frac{\partial}{\partial \gamma^2_R}\,\right\}\,N^{MFA}\Lb (\gamma^1_R|Y - Y' \Rb\,
N^{MFA}\Lb ( \gamma^2_R|Y' \Rb|_{\gamma^1_R\,=\,\gamma^2_R
\,=\,0}\,\,\,\nonumber \\
 &&=\,1\,\,-\,\,\sum^{\infty}_{n=1}\,\frac{\Lb -\gamma^{BA}\Rb^n}{n!}\,n!\,n!\,\,\,\frac{1}{(2\,\pi\,i)^2}\oint_{C^1_R}\,\,d \gamma^1_R\,\,\frac{Z^{MFA}(\gamma^1_R|Y - Y')}{(\gamma^1_R)^{n + 1}}
\,\,\,\oint_{C^2_R}\,\,\,d\,\gamma^2_R\,\,\frac{Z^{MFA}(\gamma^2_R|Y')}{(\gamma^2_R)^{n + 1}}\,\nonumber \\ &&=\,\frac{1}{(2\,\pi\,i)^2}\oint\,\oint\,\frac{d\,\tilde{\gamma}^1_R}{\tilde{\gamma}^1_R} \frac{d\,
\tilde{\gamma}^2_R}{\tilde{\gamma}^2_R}
 \,\left\{ 1-\exp \Lb \frac{\tilde{\gamma}^1_R\,\,\tilde{\gamma}^2_R}{\gamma^{BA} e^{ \Gamma(1 \to 2) Y }}\Rb\,\frac{\tilde{\gamma}^1_R\,\tilde{\gamma}^1_R}{\gamma^{BA}
e^{ {\Gamma(1 \to2)\,Y}}}\,
\Gamma\Lb 0,\frac{\tilde{\gamma}^1_R\,\tilde{\gamma}^2_R}{\gamma^{BA} e^{ {\Gamma(1 \to2)\,Y}}} \Rb \right\}\nonumber \\
&& \times\,\,\, \,\,
 Z^{MFA}\Lb \tilde{\gamma}^1_R \Rb Z^{MFA}\Lb \tilde{\gamma}^2_R \Rb \label{PL3}
\eea
Here we introduce new variables $\tilde{\gamma}^1_R\,=\,\gamma^1_R\,\exp\Lb \Gamma(1 \to2)\,(Y\,- \,Y') \Rb$ and
 $\tilde{\gamma}^2_R\,=\,\gamma^2_R\,\exp\Lb {\Gamma(1 \to2)\,Y'} \Rb$. In these new variables
\beq \label{PL4}
 Z^{MFA}\Lb \tilde{\gamma}^1_R \Rb\,\,=\,\,\frac{ 1 }{1\,\,+\,\,\tilde{\gamma}^1_R} \,;\,\,\,\,\,\,\,\,\,\,\,
 Z^{MFA}\Lb \tilde{\gamma}^2_R \Rb\,\,=\,\,\frac{ 1}{1\,\,+\,\,\tilde{\gamma}^2_R}
 \eeq
 Closing the integration on the poles  $ \tilde{\gamma}^1_R \,\,=\,\,- 1$ and $  \tilde{\gamma}^2_R \,\,=\,\,- 1$
we obtain the formula of \eq{MPSIEQ1}.

\subsection{The generating functional for the multiparticle production with  Pomeron loops }

\FIGURE[ht]{
\centerline{\epsfig{file=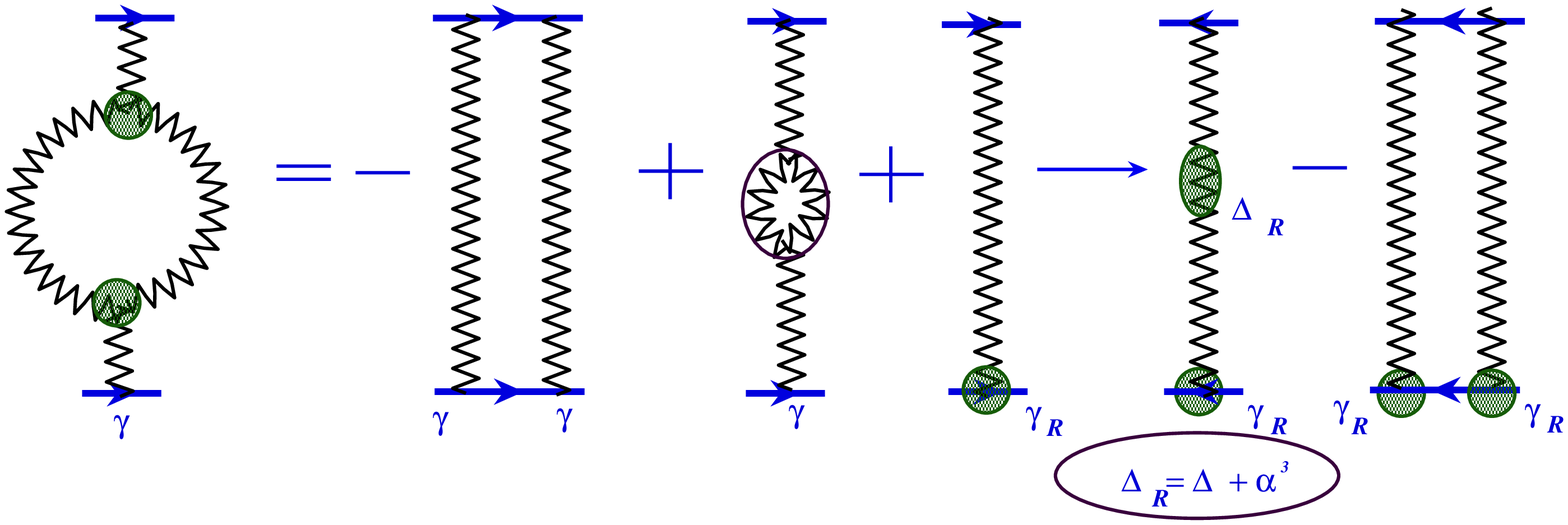,width=180mm,height=50mm}}
\caption{ The renormalization procedure  in the case of the simplest  enhanced diagram. }
\label{denre}
}

Using the result of the previous section we will derive the formula in the MPSI approach for the general
functional, defined by \eq{genw}. This formula is based on the solution of \eq{Zwv} but without the secondary derivatives. Such a solution gives the MFA approximation to our problem and we denote it as $Z^{MFA}(w,\bar{w},v| Y)$. The equation for $Z^{MFA}(w,\bar{w},v| Y)$ looks as follows
\bea \label{PL1}
&&Z\Lb  w,\bar{w},v | Y\Rb\,\,=\,\, \\
&&\frac{ w \, e^{-\Gamma(1 \to 2)  Y}}{1+w(e^{-\Gamma(1 \to 2)  Y}-1)}\,\,+\,\,
\frac{\bar{w} \, e^{-\Gamma(1 \to 2)  Y}}{1+\bar{w} (e^{-\Gamma(1 \to 2)  Y}-1)}\,\,
  -\,\,\frac{(w + \bar{w} - v) e^{-\Gamma(1 \to 2)  Y}}{1+(w + \bar{w}\,-v)(e^{-\Gamma(1 \to 2)\,\, Y}-1)} \nonumber
\eea

Using the renormalized $\gamma$ of \eq{MPSI5} we can rewrite \eq{PL1} in a different form, namely,
\bea \label{PL5}
Z^{MFA}\Lb  \gamma_R,\bar{\gamma}_R,\gamma_{in,R} | Y\Rb\,\,&=&\,\, \frac{ 1}{1+\gamma_R\,e^{\Gamma(1 \to 2)  Y}}\,\,+\,\,
\frac{1}{1+\bar{\gamma}_R \,e^{\Gamma(1 \to 2)  Y}}\,\,
  -\,\,\frac{1}{1+\xi_R\,e^{\Gamma(1 \to 2)\,\, Y}} \nonumber \\
&=&\,\, \frac{ 1}{1+\tilde{\gamma}_R}\,\,+\,\,
\frac{1}{1+\tilde{\bar{\gamma}}_R }\,\,
  -\,\,\frac{1}{1+\tilde{\xi}_R}
\eea
where we use a notation $\xi = 1 - w - \bar{w} +\,v =\,\gamma + \bar{\gamma} - \gamma_{in}$ and
\bea \label{XIR}
&\xi_R\,\,\equiv\,\,\gamma_R\,\,+\,\,\bar{\gamma}_R\,\,-\,\,\gamma_{in,R}\,\,=\,\,\,\frac{\xi}{1\,\, -\,\,\xi}\,;\,\,\,\,\,\,\,\,\,\,\,\, \xi\,\,=\,\,\frac{\xi_R}{1\,\, +\,\,\xi_R}\,;&\\
&\,\,\,\,\,\,\tilde{\gamma}_R\,=\,\gamma_R\,e^{\Gamma(1 \to 2)  Y}\,;\,\,\,\,\,\,\tilde{\bar{\gamma}}_R\,=\,\bar{\gamma}_R\,e^{\Gamma(1 \to 2)  Y}
\,;\,\,\,\,\,\,\tilde{\xi_R}\,=\,\xi_R\,e^{\Gamma(1 \to 2)\,Y} \,;&\nonumber
\eea
The first of \eq{XIR} is the definition of $\gamma_{in,R}$.

The general formula for the amplitude in the MPSI approach has the form
\bea \label{PLMPSI}
N^{MPSI}\Lb  \gamma^{BA}, \gamma^{BA}_{in}|Y\Rb&=&\left(
\exp \left\{\,-\,\gamma^{BA}\,\frac{\partial}{\partial \gamma^1_{R}}\,\frac{\partial}{\partial \gamma^2_R}
\,\,-\,\,\gamma^{BA}\,\frac{\partial}{\partial \bar{\gamma}^1_{R}}\,\frac{\partial}{\partial \bar{\gamma}^2_R}\,
\,+\,\,\gamma^{BA}_{in}\,\frac{\partial}{\partial \gamma^1_{in,R}}\,\frac{\partial}{\partial \gamma^2_{in,R}}\,\right\}
\,-\,1\right)\\
 & &Z^{MFA}\Lb  \gamma^1_R,\bar{\gamma}^1_R,\gamma^1_{in,R}|Y - Y'\Rb
  Z^{MFA}\Lb   \gamma^2_R,\bar{\gamma}^2_R,\gamma^2_{in,R}|Y'\Rb|_{\gamma^1_R\,=\,\gamma^2_R=\bar{\gamma}^1_R
=\bar{\gamma}^2_R=\gamma^1_{in,R}=\gamma^2_{in,R}\,=\,0} \nonumber \\
&=&
\left(\exp \left\{\, - \,\gamma^{BA}\,\frac{\partial}{\partial \tilde{\gamma}^1_{R}}\,\frac{\partial}{\partial\tilde{\gamma}^2_R}
- \gamma^{BA}\,\frac{\partial}{\partial\tilde{\bar{\gamma}}^1_{R}}\,\frac{\partial}{\partial\tilde{\bar{\gamma}}^2_R}
 +  \gamma^{BA}_{in}\,\frac{\partial}{\partial\tilde{\gamma}^1_{in,R}}\,\frac{\partial}{\partial \tilde{\gamma}^2_{in,R}}\,\right\}\,-\,1 \right)\nonumber \\
 & &Z^{MFA}\Lb  \tilde{\gamma}^1_R,\tilde{\bar{\gamma}}^1_R,\tilde{\gamma}^1_{in,R}\Rb
 Z^{MFA}\Lb \tilde{\gamma}^2_R,\tilde{\bar{\gamma}}^2_R,\tilde{\gamma}^2_{in,R}  \Rb|_{\gamma^1_R\,=\,\gamma^2_R=\bar{\gamma}^1_R
=\bar{\gamma}^2_R=\gamma^1_{in,R}=\gamma^2_{in,R}\,=\,0} \nonumber
\eea
where $\gamma^{BA}$ and $\gamma^{BA}_{in}$ are the elastic and inelastic  amplitudes of interaction of two dipoles
 at low energy which are calculated in QCD in the Born approximation. The plus sign in \eq{PLMPSI} in front of $
   \gamma^{BA}_{in}\,\frac{\partial}{\partial\tilde{\gamma}^1_{in,R}}\,\frac{\partial}{\partial \tilde{\gamma}^2_{in,R}}$ reflects the fact that the Pomeron loop with two cut Pomerons does not have a negative contribution unlike in the case of uncut Pomerons. The sign between the exponent and unity in \eq{PLMPSI} could be easily checked  noticing the first term of the expansion of the exponent correctly reproduces the positive contribution of $\gamma_{in}$ ( cut Pomeron).

 The nice feature of this equation that one can see that the result does not depend on the value of  an arbitrary chosen  rapidity $Y'$. Using the explicit form for $Z^{MFA}$  given
by \eq{PL5} we can calculate $Z^{MPSI}$ in closed form denoting by
\beq \label{DEFG}
G\Lb x \Rb\,\,\equiv\,\,
\exp \Lb \frac{1}{x}\Rb\,\frac{1}{x}\,\,
\Gamma\Lb 0,\frac{1}{x} \Rb
\eeq
$N^{MPSI}$  is equal to
\beq \label{PLMPSI1}
N^{MPSI}\Lb  \gamma^{BA}, \gamma^{BA}_{in}|Y\Rb\,\,= \,\,2\,\Lb 1 \,-\,G\Lb \gamma^{BA}\,
e^{\Gamma(1 \to 2)\,Y} \Rb \Rb\,-\,\Lb 1 \,-\,G\Lb \Lb 2\,\gamma^{BA}\,-\,\,\gamma^{BA}_{in} \,\Rb\,e^{\Gamma(1 \to 2)\,Y} \Rb \Rb
\eeq

The useful formulae for getting \eq{PLMPSI1}  are  the following
\beq \label{PL6} \frac{\partial^k}{\partial\,\gamma^{k}_{in}}\,
\frac{\partial^{l_1}}{\partial\,\gamma^{l_1}}
\frac{\partial^{l_2}}{\partial\,\bar{\gamma}^{l_2}}\,\,\frac{1}{ 1 +
\gamma + \bar{\gamma}\,-\,\gamma_{in}} \,\,\,=\,\,(- 1)^{l_1 +
l_2}\,(l_1 + l_2 + k)! \eeq and    equations  {\bf 8.350 - 8.359}
for the incomplete gamma function $\Gamma \Lb 0,x \Rb$ in
\cite{RY}.

\eq{PLMPSI1} allows us to calculate the cross section with fixed multiplicity of produced particles. Namely, the cross section for the processes with $k\,<n>$ particles in the final state, where $<n>$ is the mean multiplicity in our  reaction,
can be calculated as
\beq \label{PL7}
\sigma_k\,(Y)\,\,=\,\frac{1}{k!}\,\,\,\Lb\frac{\partial^k}{\partial (\gamma^{BA}_{in})^k }\,\,N^{MPSI}\Lb  \gamma^{BA}, \gamma^{BA}_{in}|Y\Rb\Rb|_{\gamma^{BA}_{in} = 0}\,\,\cdot\,\Lb\gamma^{BA}_{in} = 2 \gamma^{BA}\Rb^k\,
\eeq
Here we use that $\gamma^{BA}_{in}\,\,=\,\,2\,\gamma^{BA}$  in the Born approximation of QCD.

It is interesting to check the general equation (see \eq{PLMPSI1})   calculating two known cases: the diffractive dissociation process and the total inelastic  cross section. The first one can be calculated using  \eq{PLMPSI1}   with $\gamma^{BA}_{in}\,=\,0$.
 The answer is
  \beq
\label{PL8}
N^{MPSI}_{diff}\,\Lb \gamma^{BA}|Y\,\Rb\,\,=\,\,2\,N^{MPSI}_0\Lb \gamma^{BA}|Y\,\Rb\,\,-\,\,N^{MPSI}_0\Lb 2\, \gamma^{BA}|Y\,\Rb
\eeq
where $N_0$ is given by \eq{MPSIEQ1}. \eq{PL8} is a direct consequence of the unitarity constraints( see \eq{FUC}). As you can see from \eq{SolZ} the same formula determines the diffractive production in the mean field approximation.

The value of the total inelastic cross section, that is equal the sum of diffractive production and inelastic cross section,  stems from \eq{PLMPSI1}   for $\gamma^{BA}_{in} = 2\,\gamma^{BA}$
and it is equal to
\beq
\label{PL9}
N^{MPSI}_{total\,\,\, inelastic}\,\Lb \gamma^{BA}|Y\,\Rb\,\,=\,\,2\,N^{MPSI}_0\Lb \gamma^{BA}|Y\,\Rb
\,=\,N^{MPSI}_{diff}\,\Lb \gamma^{BA}|Y\,\Rb\,\,+\,\,N^{MPSI}_{inel}\,\Lb \gamma^{BA}|Y\,\Rb
\eeq
which is actually the unitarity constraint itself ( see \eq{FUC}). One can see that the inelastic cross section is determined by
\beq \label{PL10}
N^{MPSI}_{inel}\,\Lb \gamma^{BA}|Y\,\Rb\,\,=\,\, 1 \,\,-\,G\Lb\,\,2\,\gamma^{BA}\,e^{\Gamma(1 \to 2)\,Y} \Rb
\eeq

\section{Conclusions}
In this paper we introduce a new generating function for the processes of multiparticle production both for the mean field approximation (see \eq{NZ} and for the general case (see \eq{genw}). For general case where the Pomeron loops
has been taken into account, we obtain the linear evolution equation for the generating function (see \eq{Zwv})
while in the mean field approximation we proved both the linear evolution equation (see \eq{Zuv}) and the non-linear equation (see \eq{NonLin2}). The last one is the generalization of Kovchegov-Levin equation for diffractive production to a general case of the processes with arbitrary multiplicities. Since this equation is proven for the general QCD case we hope that the equations for the general  generating function can be proven for the real QCD evolution.

The second result of the paper is the new method of summing the Pomeron loops. We argued that the sum of all Pomeron diagrams, including loops, in the kinematic region of \eq{KR} can be reduced to the diagrams of the Pomeron exchanges without interactions between Pomeron if we renormalize the amplitude of low energy interaction. Based on this result we suggest a generalization of the Mueller-Patel-Salam-Iancu method of summation of
the Pomeron loops. In particular, we calculated the new generating function for the inelastic processes in the improved MPSI approximation
( see  \eq{PLMPSI1}).

We would like to stress that we firmly believe that the scattering amplitude, calculated using this method, leads to a correct answer to the old problem: the high energy asymptotic behaviour of the scattering amplitude at ultra high energies beyond of the  BFKL Pomeron calculus in kinematic region of \eq{KR}.

We hope that both results will lead to  new simplifications in the
case the BFKL Pomeron calculus in QCD (in two transverse
dimensions). The general case of the BFKL Pomeron calculus in QCD
will be addressed in a separate paper. We would like also to mention
that this case has been started to discuss in   \cite{KLW}.

\section*{Acknowledgments:}
We are very grateful to  Jochen Bartels, Asher Gotsman, Edmond Iancu,  Michael Kozlov, Uri Maor and Al Mueller
for fruitful discussions on the subject.
Special thanks of A.P. goes to Jochen Bartels for his hospitality and support during  his visit to DESY.
This research was supported in part  by the Israel Science Foundation,
founded by the Israeli Academy of Science and Humanities and by BSF grant \# 20004019.

We thank the Galileo Galilei Institute for Theoretical
Physics for the hospitality and the INFN for partial support during the
completion of this work.

\end{document}